\documentclass[%
 reprint,
 superscriptaddress,
 amsmath,amssymb,
 aps,
 prx,
]{revtex4-2}
\usepackage{graphicx}
\usepackage{bbold}
\usepackage{tikz}
\usepackage{subfig}
\usepackage[many]{tcolorbox}
\usepackage{url}
\newtcolorbox{cross}{blank,breakable,parbox=false,
  overlay={\draw[red,line width=5pt] (interior.south west)--(interior.north east);
    \draw[red,line width=5pt] (interior.north west)--(interior.south east);}}
\usepackage{dsfont}
\usepackage{amsfonts}
\usepackage{amsthm}
\usepackage{braket}
\usepackage{amsmath}
\usepackage{algorithm}
\usepackage{algpseudocode}
\usepackage{comment}
\theoremstyle{definition}
\newtheorem{definition}{Definition}[section]
\usepackage{dcolumn}
\usepackage{bm}

\usepackage{algorithm}
\usepackage{algpseudocode}

\begin{document}

\preprint{APS/123-QED}

\title{Beating classical heuristics for the binary paint shop problem with the quantum approximate optimization algorithm}

\author{Michael Streif}
\affiliation{Data:Lab, Volkswagen Group, Ungererstr. 69, 80805 München, Germany}
\affiliation{University Erlangen-Nürnberg (FAU), Institute of Theoretical Physics, Staudtstr. 7, 91058 Erlangen, Germany}

\author{Sheir Yarkoni}
\affiliation{Data:Lab, Volkswagen Group, Ungererstr. 69, 80805 München, Germany}
\affiliation{LIACS, Leiden University, Snellius Gebouw, Niels Bohrweg 1, 2333 CA Leiden, Netherlands}

\author{Andrea Skolik}
\affiliation{Data:Lab, Volkswagen Group, Ungererstr. 69, 80805 München, Germany}
\affiliation{LIACS, Leiden University, Snellius Gebouw, Niels Bohrweg 1, 2333 CA Leiden, Netherlands}

\author{Florian Neukart}
\affiliation{Data:Lab, Volkswagen Group, Ungererstr. 69, 80805 München, Germany}
\affiliation{LIACS, Leiden University, Snellius Gebouw, Niels Bohrweg 1, 2333 CA Leiden, Netherlands}

\author{Martin Leib}
\affiliation{Data:Lab, Volkswagen Group, Ungererstr. 69, 80805 München, Germany}

\date{\today}

\begin{abstract}
The binary paint shop problem (BPSP) is an APX-hard optimization problem of the automotive industry. In this work, we show how to use the Quantum Approximate Optimization Algorithm (QAOA) to find solutions of the BPSP. We demonstrate that QAOA with constant depth is able to beat all known heuristics for the binary paint shop problem on average in the infinite size limit $n\rightarrow\infty$. We complete our studies by running first experiments of small-sized instances on a trapped-ion quantum computer through Amazon Braket. 
\end{abstract}

\maketitle

\newpage
\section{Introduction}
Achievements in fabrication and control of two-level systems led to first quantum computers with tens of qubits \cite{barends2016digitized,dicarlo2009demonstration,debnath2016demonstration,reagor2018demonstration} and recently culminated in the demonstration of a quantum computer  solving a computational task intractable for classical computers, also known as \textit{quantum supremacy} \cite{arute2019quantum}. This milestone raises expectations that quantum computing some day will accelerate research, speed up simulations in chemistry and improve optimization processes in many branches of industry. Quantum algorithms with proven scaling advantage over classical algorithms, such as Grover's \cite{grover1996fast} or Shor's \cite{shor1994algorithms} algorithm, require fault-tolerant quantum computers. However the devices which will become available in the next 5 to 10 years will only have a limited number of qubits and will not feature error-correction. It is unclear if such Noisy Intermediate-Scale Quantum (NISQ) devices can be useful in solving real-world problems faster than classical computers or if larger error-corrected devices will be needed. To answer this question it is especially important to develop quantum algorithms tailor-made to the quantum processing units' (QPU) characteristics. A promising class of NISQ algorithms is the class of variational quantum algorithms, which are parameterized ansätze optimized with classical learning loops. There exist various ideas how to tailor the ansatz for different tasks, such as the Variational Quantum Eigensolver (VQE) for chemistry applications \cite{peruzzo2014variational} or Quantum Neural Networks (QNN) for machine learning \cite{mitarai2018quantum,killoran2019continuous}. The QAOA is a variational algorithm designed to solve classical optimization problems \cite{farhi2014quantum} and was applied to problems such as Max-Cut \cite{farhi2014quantum} or Max-3-Lin-2 \cite{2014arXiv1412.6062F}. Furthermore, there exist first insights on QAOAs performance \cite{crooks2018performance, streif2019comparison, streif2020forbidden,willsch2020benchmarking}, first experimental realizations on different quantum processors \cite{arute2020quantum, pagano2020quantum, otterbach2017unsupervised} and several proposals how to further improve QAOA \cite{zhou2020quantum,  mbeng2019quantum, bapat2018bang, hastings2019classical, bravyi2019classical, brandao2018fixed, Streif_2020}. For example in \cite{Streif_2020}, it was shown that for some problem classes with certain topological characteristics it is possible to find good parameters for QAOA with classical methods efficiently.  Moreover, there exist results showing that it is classically hard to sample from the QAOA output \cite{farhi2016quantum} and that QAOA possesses a Grover-type speed-up \cite{jiang2017near}. However, performance bounds are only known for very short circuits \cite{farhi2014quantum} or classically easy instances \cite{mbeng2019quantum}. Establishing scaling comparisons, beyond low depth circuits, between QAOA and classical methods for application relevant optimization  problems is the next important step to find useful NISQ applications of QAOA.

In this present contribution, we take a step in this direction using the example of a combinatorial optimization problem from the automotive industry, the \textit{binary paint shop problem} (BPSP). We show that the problem can be encoded into a spin glass in a constraint-free way and requires only linear number of qubits for increasing problem size. As such it is a perfect fit for QAOA on NISQ devices. We show that the problem has fixed degree and coupling strength, which allows us to use the method developed in \cite{Streif_2020} which bypasses the NP-hard training procedure \cite{bittel2021training}. We present numerical results and run first small-scale experiments on a trapped ion quantum computer. We are numerically able to prove a minimal depth for QAOA to beat classical heuristics in the small to medium system size limit (up to 100 cars) as well as for the infinite size limit. With this we show that a constant time quantum algorithm can beat a polynomial time classical algorithm.

This paper is structured as follows. In Sec.~\ref{sec:qaoa} we review the Quantum Approximate Optimization Algorithm (QAOA). In Sec.~\ref{sec:bpsp} we review the binary paint shop problem, classical greedy algorithms to solve it and discuss the mapping of the problem onto an Ising Hamiltonian. In Sec.~\ref{sec:solving_the_bpsp_with_qaoa} we show results of QAOA applied to the BPSP and in Sec.~\ref{sec:conclusion}, we conclude. 

\section{Review of QAOA}
\label{sec:qaoa}
In this section, we review the Quantum Approximate Optimization Algorithm (QAOA) \cite{farhi2014quantum}, a variational quantum algorithm designed to solve combinatorial optimization problems. 

To solve a combinatorial optimization problem with QAOA, the first step is to reformulate it as a spin glass problem. For our purposes the spin glass can be represented as a problem graph $G=(V,E)$ with nodes $v\in V$ representing spins $s_i$ and edges $e\in E$ representing the terms of the sum of the energy of the spin glass $E_\mathrm{P} = \sum_{(i,j) \in E} J_{i,j} s_i s_j$ that needs to be minimized. We note that finding the optimal solution of a spin glass is $\mathcal{NP}$-hard \cite{barahona1982computational}, thus there exist mappings with at most polynomial overhead from all $\mathcal{NP}$ problems to such a spin system, some of them shown in \cite{lucas2014ising}. We search for low energy configurations of the spin glass with a variational ansatz. In QAOA, this is done by minimizing the expectation value of the problem Hamiltonian $H_\mathrm{P}$  with respect to a parameterized ansatz state $\ket{\Psi(\{\beta_l, \gamma_l\})}$,
\begin{align}
    \label{eq:expectation_value_qaoa}
    \min_{\left\{\beta_l, \gamma_l\right\}} \left\langle \Psi(\{\beta_l, \gamma_l\})| H_{\mathrm{P}}| \Psi(\{\beta_l, \gamma_l\})\right\rangle \,.
\end{align}
Therefore we translate the spin system to its quantum version
\begin{align}
    \label{eq:cost_hamiltonian}
    \mathrm{H}_\mathrm{P}=\frac{1}{2}\sum_{(i,j) \in E}J_{ij}\sigma_z^{(i)}\sigma_z^{(j)}\,,
\end{align}
where each classical spin variable $i$ is replaced by a qubit $i$ with the Pauli-Z operator $\sigma_z^{(i)}$. 
 The ansatz state in QAOA is inspired by quantum annealing techniques and is generated by the repeated application of the mixing and problem unitary, $U_{\mathrm{M}}(\beta_l) = e^{-i\beta_l H_\mathrm{M}}$ and $U_{\mathrm{P}}(\gamma_l)=e^{-i\gamma_l H_\mathrm{P}}$, on the superposition state of all computational basis states, $\left|+\right\rangle^{\otimes n}=\bigotimes_i^n 1/\sqrt{2}\left(\ket{0}_i+\ket{1}_i\right)$. The generators of these unitaries are given by the mixing Hamiltonian, $H_\mathrm{M}=\sum_i \sigma_x^{(i)}$, and the problem Hamiltonian, see Eq.~(\ref{eq:cost_hamiltonian}). The full ansatz state 
\begin{align}
    \label{eq:qaoa_ansatz}
    \left|\Psi(\{\beta_l, \gamma_l\})\right\rangle_\mathrm{QAOA} = \prod_l^{p} U_\mathrm{X}(\beta_l)U_\mathrm{P}(\gamma_l) \left|+\right\rangle^{\otimes n}\,,
\end{align}
includes $p$ repetition of those unitaries, where each repetition is called a QAOA level. To find the optimal variational parameters $\{\beta_l^*, \gamma_l^*\}_{l=1}^p$, a quantum computer is used to estimate the expectation value of the problem Hamiltonian, while an outer learning loop on a classical computer updates the parameters to minimize the expectation value. Recent work has shown how to speed up the classical learning loop \cite{zhou2020quantum} or alternatively using entirely classical methods to find optimal parameters for certain problem classes \cite{Streif_2020}. Having found the optimal parameters, one then samples from the final state $\ket{\Psi(\{\beta_l^*, \gamma_l^*\})}$ in the computational basis which yields solutions to the optimization problem. 

\section{The binary paint shop problem}
\label{sec:bpsp}
\begin{figure}[t!]
    \centering
 \includegraphics[width=1\linewidth]{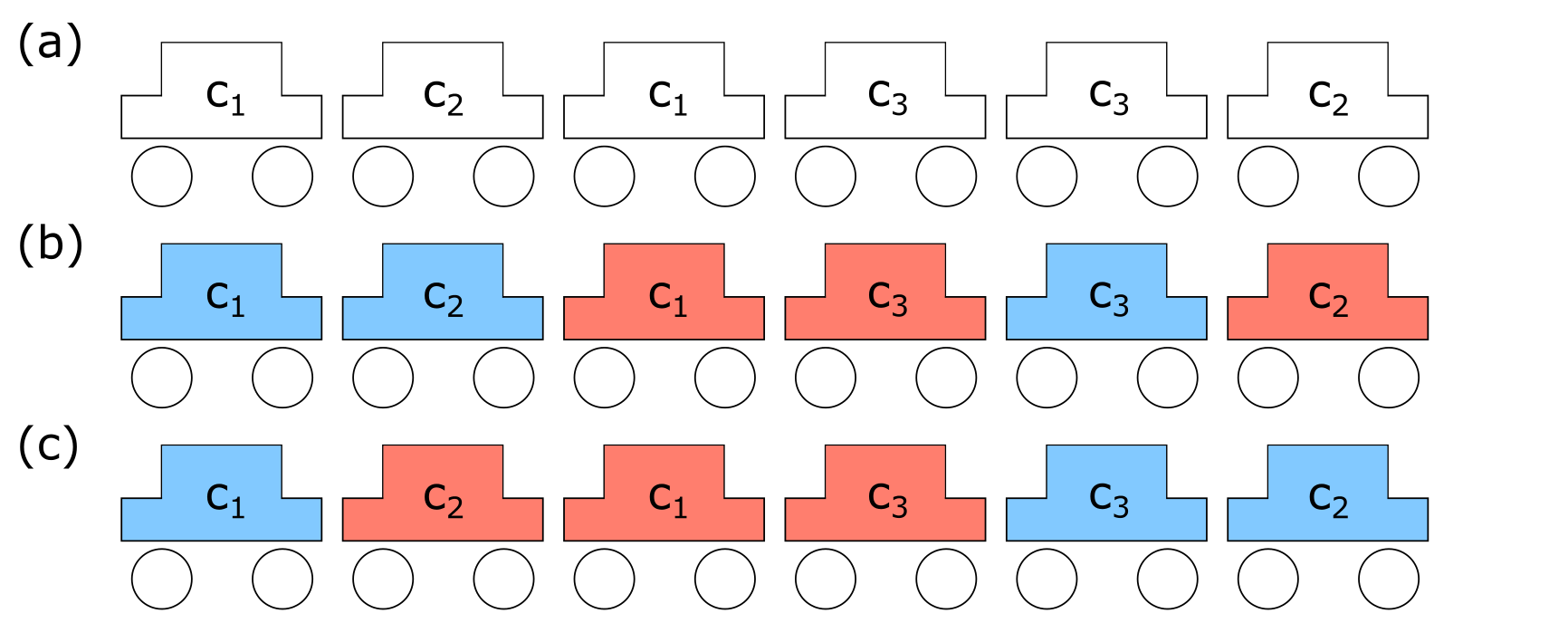}
  \caption{(a) A binary paint shop instance with $n=3$ cars $\{c_1,c_2,c_3\}$. (b) A valid but sub-optimal coloring with $\Delta_C=3$ color changes. (c) An optimal coloring which only requires $\Delta_C=2$ color changes to paint the sequence. }
\label{fig:bpsp}
\end{figure}

Assume an automotive paint shop and a random, but fixed sequence of $n$ cars. The task is to paint the cars in the order given by the sequence. Each individual car needs to be painted with two colors, once per color, in a to-be-determined color order. Meaning, each car appears twice at random, uncorrelated positions in the sequence and we are free to choose the color to paint the car first. A specific choice of first colors for every car is called a \emph{coloring}. The objective of the optimization problem is to find a coloring which minimizes the number of color changes between adjacent cars in the sequence. This combinatorial optimization problem is called the \textit{binary paint shop problem} (BPSP) \cite{epping2001some,epping2004complexity,gupta2013approximability}. In Fig.~\ref{fig:bpsp}, we show a binary paint shop instance together with a sub-optimal and the optimal solution. A formal definition of the binary paint shop problem is given in Def.~\ref{def:bpsp}.

\theoremstyle{definition}
\begin{definition}{(Binary paint shop problem)}
\label{def:bpsp}
Let $\Omega$ be the set of $n$ cars $\{c_1,\dots,c_n\}$. An instance of the \textit{binary paint shop problem} is given by a sequence $\left(w_1,\dots,w_{2n}\right)$ with $w_i\in \Omega$, where each car $c_i$ appears exactly twice. We are given two colors $\mathcal{C}=\{1,2\}$. A coloring is a sequence $f={f_1,\dots,f_{2n}}$ with $f_i\in\mathcal{C}$ and $f_i\neq f_j$ if $w_i=w_j$ for $i\neq j$. The objective is to minimize the number of color changes $\Delta_C = \sum_i |f_i-f_{i+1}|$.
\end{definition}

The binary paint shop problem belongs to the class of $\mathcal{NP}$-hard optimization problems, thus there is no polynomial-time algorithm which finds the optimal solution for all problem instances. For many optimization problems in practice, rather than spending exponential time to find the optimal solution, fast approximate algorithms are used.
However, the binary paint shop problem is proven to be $\mathcal{APX}$-hard \cite{epping2004complexity}, i.e. it is as difficult to approximate as every problem in $\mathcal{APX}$. Additionally, if the Unique Games Conjecture (UGC) \cite{khot2002power} holds, it is even not in $\mathcal{APX}$ and thus, there would not be a constant factor approximation algorithm for any $\alpha$ \cite{gupta2013approximability}. A constant factor approximation algorithm would be a polynomial-time algorithm which returns an approximate solution with at most $\alpha \mathrm{OPT}$ color changes where $\mathrm{OPT}$ is the optimal number of color changes.  This is a key difference to previous problem classes QAOA has been applied to, such as Max-Cut, where constant factor approximation algorithms are known \cite{goemans1995improved}. 

Several greedy algorithms exist for the binary paint shop problem which provide solutions with color changes linear in the number of cars $n$ on average \cite{epping2004complexity, epping2001some}. The greedy algorithm introduced in \cite{epping2001some} starts at the first car $w_1$ of the sequence with one of both colors, goes through the sequence and changes colors when necessary, i.e. only if the same car would be painted twice with the same color, see Fig.~\ref{alg:greedy} for pseudo code of this algorithm. For $n\rightarrow\infty$ cars, this greedy algorithm finds a solution with an average number of color changes $\mathbb{E}_\mathrm{G}(\Delta_C)=n/2$  \cite{andres2011some}. In Appendix \ref{app:classicalheuristics}, we review two other greedy algorithms, the red first algorithm and the recursive greedy algorithm, yielding $\mathbb{E}_\mathrm{RF}(\Delta_C)=2n/3 $ and $\mathbb{E}_\mathrm{RG}(\Delta_C)= 2n/5$ color changes on average respectively \cite{andres2011some}. Numerical results however suggest that the average number of optimal color changes is sub-linear in the number of cars $n$ \cite{meunier2012computing}. Moreover, for some instances, the greedy algorithms only find solutions far from the optimal solution \cite{andres2011some}. For example, for the instance shown in Fig.~\ref{fig:bpsp}(a), the greedy algorithm finds the solution given in (b) rather then the optimal solution shown in (c).

More general versions of the binary paint shop problem can be found in practice. Typically both the color set is augmented to contain more than two colors, and identical cars appear more than twice per word. These conditions correspond to the real-world application of painting thousands of car bodies per day, with numerous colors. Even restricting the color set to two colors has real-world relevance: before painting the final color of the body, each car is first painted with an undercoat called a \emph{filler} coat. The filler colors are restricted to white and black, depending on the final car body color, and optimizing the number of color switches within the filler queue yields production cost savings. Given that the generalized paint shop problem is $\mathcal{NP}$-hard in both the number of cars and colors~\cite{epping2004complexity}, and that the binary color set is already industrially relevant, we restrict the color set to two colors in this study.   

\begin{figure}
\begin{algorithm}[H] 
\caption{Greedy algorithm}
\begin{algorithmic}[1]
\Statex \textbf{Input}:  a sequence $(w_{1} \dots w_{2n})$, two colors $\mathcal{C}=\{1,2\}$
\Statex \textbf{Output}:  a coloring $f$
\Statex
\Function{greedy}{w}
    \State first color of each car $c_i$: $\mathrm{FC}(c_i)=\mathrm{None} \phantom{.}\forall i$
    \State Choose one of both colors: $c=1,2$

    \State
    \For{$k \gets 1$ to $length(w)$}
        \State $car \gets w_k$
        \If{$\mathrm{FC}(car)\neq \mathrm{None}$}
            \State $f_k= c$
            \State $\mathrm{FC}(car)=c$
        \Else
            \State $c=(c+1)\mod 2$
            \State $f_k = c$
        \EndIf
    \EndFor
    \State \Return {coloring sequence $f$}
\EndFunction

\end{algorithmic}
\end{algorithm}
\caption{Pseudo code of the greedy algorithm to solve a binary paint shop instance.}
\label{alg:greedy}
\end{figure}

\subsection{Reformulating the BPSP as a spin glass}
\label{sec:mapping_to_spins}
In this section, we explain how to map the binary paint shop problem to a problem Hamiltonian as in Eq.~(\ref{eq:cost_hamiltonian}). 

We start by assigning a single qubit $i$ to each car $c_i$ in the sequence. The eigenstates of the $\sigma_z$-operator of each qubit indicate in which color each car is painted first. The second color choice for each car in the sequence is then unambiguously determined. To penalize color changes in the coloring, we use the coupling strengths $J_{ij}$ between the qubits. We start at the first car $w_1$ in the sequence and step through the sequence adding couplings between the qubits representing the car $w_k$ and its next neighbor $w_{k+1}$ in the sequence. If both cars $c_i$ and $c_j$, represented by $w_k$ and $w_{k+1}$ respectively, appear both for the first or second time, a ferromagnetic coupling, $J_{ij}=-1$, is added. This ensures that consecutive cars favor to be painted with the same color. If either car has already appeared in the sequence while the other has not, we instead add an antiferromagnetic coupling, $J_{ij}=1$. We know that a solution with $\Delta_C$ color changes is separated by an energy $\Delta E=1$ from a solution with $\Delta_C+1$ color changes. We show pseudo code of this mapping in Fig.~\ref{alg:spin_glass_mapping}. 
We note that the encoding of the problem does not include any constraints, thus all computational states embody valid solutions to the problem. Moreover, the encoding only requires $n$ qubits for $n$ cars, making it a better fit for NISQ devices than typical scheduling problems (like the traveling salesman problem) where the number of qubits required grows quadratically with the system size \cite{lucas2014ising}. From the BPSP construction, we also know that a solution with $\Delta_C$ color changes is separated by an energy $\Delta E=1$ from a solution with $\Delta_C+1$ color changes. 

\begin{figure}
\begin{algorithm}[H] 
\caption{Mapping of the binary paint shop problem onto a spin glass}
\begin{algorithmic}[1]
\Statex \textbf{Input}: a sequence representing a BPSP instance, cf. Def.~\ref{def:bpsp}
\Statex \textbf{Output}: an Ising Hamiltonian $H_\mathrm{P}$
\Statex
\Function{mapping}{w}
  \State $H_\mathrm{P}=0$
  \State associate car $c_i$ with qubit $i$
  \State \#$c_i=0 \phantom{.}\forall i$
    \For{$k \gets 1$ to $length(w)$}
        \State $car_1,car_2 \gets w_k, w_{k+1}$
        \State $H_\mathrm{P}\gets H_\mathrm{P} + (-1)^{\#car_1+\#car_2+1}\sigma_z^{(car_1)}\sigma_z^{(car_2)}$
        \State $\#car_1 \gets \#car_1+1$
    \EndFor
    \State \Return {$\mathrm{H}_\mathrm{C}$}
\EndFunction
\end{algorithmic}
\end{algorithm}
\caption{Pseudo code for mapping a binary paint shop instance with $n$ cars to an Ising Hamiltonian with $n$ qubits.}
\label{alg:spin_glass_mapping}
\end{figure}
\subsubsection{Properties of the Ising Hamiltonian}
\label{sec:properties_spins}

In \cite{Streif_2020} it was shown how to calculate close-to-optimal QAOA parameters using a classical computer $\left\{\beta_i^\mathrm{tree}, \gamma_i^\mathrm{tree}\right\}$ for various levels $p$, and problem classes represented by graphs with fixed degree and uniform coupling strengths, $J_{ij}=const.$. For the NISQ era, where typically $p$ is small, this method circumvents optimizing the variational parameters $\left\{\beta_i, \gamma_i\right\}$ of the QAOA ansatz state for each instance independently, and thus reduces the total QPU time used. In this section, we show that the binary paint shop instances represent graphs of fixed and average degree $4$ and coupling strengths $J_{ij}=\pm 1$ (both for $N\rightarrow\infty$). As this method originally was proposed for the case $J_{ij}=J$ only, we prove in Appendix \ref{app:tree-qaoa} that the method also works if $|J_{ij}|=const.$. In the following we argue that the BPSP is a perfect fit for parameters calculated using this method.

\paragraph{Average degree}
In the construction of the problem Hamiltonian, cf. Sec.~\ref{sec:mapping_to_spins}, we add an interaction between two qubits if the corresponding cars are adjacent. As each car only appears $2$ times, each car has maximal $4$ distinct neighbors. It follows that the nodes in the graph $G$ representing the spin system also have maximal degree of $4$. The degree is only smaller than $4$ for the node representing the first or the last car in the sequence, or if the car is adjacent to the same car twice. In Fig.~\ref{fig:coupling_prob_and_average_degree}(a), we show that the average degree of the graph is converging to $4$ from below and becomes effectively $4$-regular when $n\rightarrow\infty$.

\paragraph{Coupling strengths $J_{ij}$}
From the construction of the Ising Hamiltonian, we also know that the interaction values $J_{ij}$ are integers and given by $\{-2,-1,1\}$. In Appendix~\ref{app:coupling_strengs}, we show that the distribution of interaction values $P(J_{ij})$ converges to a distribution with $P(J_{ij}=-1)=\frac{2}{3}$ and $P(J_{ij}=1)=\frac{1}{3}$, when $n\rightarrow\infty$. This means that the probability of a ferromagnetic coupling is twice as big as for an anti-ferromagnetic coupling and that coupling strengths $|J_{ij}|=2$ are suppressed in the infinite size limit. We note however that the coupling strengths $J_{ij}$ are not independent from each other. 

\section{Solving the Binary Paint Shop problem with QAOA}
\label{sec:solving_the_bpsp_with_qaoa}
In the following sections, we apply QAOA to the binary paint shop problem.  For all simulations and experiments, we use the parameters $\left\{\beta_i^\mathrm{tree}, \gamma_i^\mathrm{tree}\right\}$ found with method \cite{Streif_2020}, shown in Tab.~\ref{tab:tree-qaoa-parameters}.

\subsection{Numerical results}
\label{sec:numerics}
\begin{figure}[t!]
    \centering
 \includegraphics[width=1\linewidth]{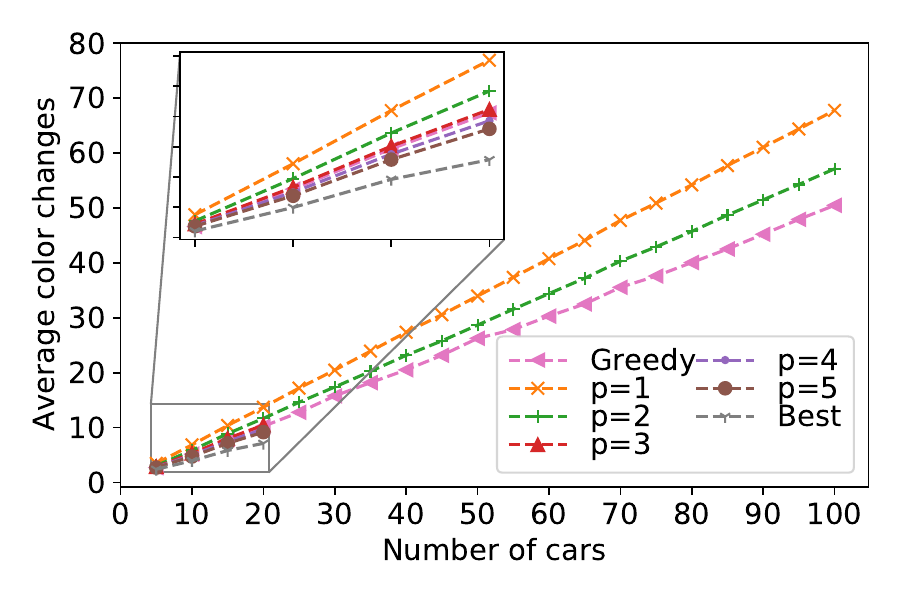}

  \caption{Numerical results for the binary paint shop problem. The classical greedy algorithm is compared to QAOA with different levels $p$. Each data point is averaged over $100$ randomly generated instances.}
\label{fig:numerics}
\end{figure}
In this section, we numerically analyze the performance of QAOA on $100$ randomly generated binary paint shop instances of sizes from $n=5$ to $n=100$ cars in increments of $5$ cars \cite{bpsp_data}.

For up to $n=20$ cars we calculate the QAOA output state, Eq.~(\ref{eq:qaoa_ansatz}) for $p\in\{1,2,3,4,5\}$ levels of QAOA and determine the energy expectation value Eq.~(\ref{eq:expectation_value_qaoa}) together with the expected number of color changes $\Delta_{C}$. For larger systems, the calculation of the QAOA output state is out of reach using a standard desktop computer.

However, since we are only interested in the energy expectation value and not in the output state, we use a proxy to calculate the expectation value for small values of $p$. First, we rewrite Eq.~(\ref{eq:expectation_value_qaoa}) as a sum over all expectation values of two-point correlation functions,
\begin{align}
    \label{eq:sum_expectation_value}
    \braket{H_\mathrm{P}}=\sum_{i,j} \frac{1}{2}\braket{\sigma_z^{(i)}\sigma_z^{(j)}}\;.
\end{align}
As pointed out in \cite{farhi2014quantum, Streif_2020}, the individual expectation values $\braket{\sigma_z^{(i)}\sigma_z^{(j)}}$ do not necessarily depend on the states of all qubits, but only on a subset, which can be used to reduce the computational cost to calculate them. Some of the gates defined in $U^{\textrm{QAOA}} = \prod_i^{p} U_\mathrm{X}(\beta_i)U_\mathrm{P}(\gamma_i)$ commute with the operator product $\sigma_z^{(i)}\sigma_z^{(j)}$, and since the gates are unitary we can completely skip them in the definition of the correlation function,
\begin{multline}
    \braket{\sigma_z^{(i)}\sigma_z^{(j)}} = \\ \bra{+_{\text{supp}(\text{RCC}_{i,j})}} U^{\textrm{QAOA}\dag}_{\textrm{RCC}_{i,j}} \sigma_z^{(i)}\sigma_z^{(j)} U^{\textrm{QAOA}}_{\textrm{RCC}_{i,j}} \ket{+_{\text{supp}(\text{RCC}_{i,j})}}\,,
\end{multline}
where $\text{RCC}_{i,j}$ is the set of gates not commuting with $\sigma_z^{(i)}\sigma_z^{(j)}$ called the reverse causal cone of the correlation function, $\text{supp}(\text{RCC}_{i,j})$ is the support of the reverse causal cone, i.e. the minimal set of qubits the reverse causal cone acts on, and $\ket{+_{\text{supp}(\text{RCC}_{i,j})}}=\bigotimes_{l \in \text{supp}(\text{RCC}_{i,j})} 1/\sqrt{2}\left(\ket{0}_l+\ket{1}_l\right)$ is the superposition state of all computational basis states of the qubits in the reverse causal cone.
The support of the reverse causal cone can be constructed in an iterative procedure \cite{farhi2014quantum, Streif_2020}: for each layer in the QAOA circuit we add all new neighbors in the problem graph of the support of the reverse causal cone of a QAOA circuit with one level less starting with the two qubits that define the correlation function. Therefore, the number of qubits affecting the expectation value depends on the number of QAOA levels $p$ and the topology of the problem. The binary paint shop instances can be represented as graphs with bounded degree of $4$, thus the reverse causal cone includes up to $3^{p+1}-1$ qubits. For $p=\{1,2\}$ this results in system sizes that can be simulated using a standard desktop computer, independent of the actual size of the instance. After calculating the individual correlation functions independently we find the QAOA expectation value by using Eq.~(\ref{eq:sum_expectation_value}).

In Fig.~\ref{fig:numerics}, we show the expected color changes from the QAOA output averaged over all instances together with the average result of the greedy algorithm, see Fig.~\ref{alg:greedy}, and exact solutions for up to $n=20$ cars. Low-depth QAOA with $p=\{1,2\}$ performs worse than the polynomial-time greedy algorithm, while for $p=3$ levels the performance gap nearly vanishes. For $p=\{4,5\}$ QAOA outperforms the greedy algorithm. 
 
\subsection{Beating the greedy algorithms for large instances}
The greedy algorithms presented in Sec.~\ref{sec:bpsp} provide solutions with color changes growing linearly with the number of cars on average in polynomial time. Thus, they provide a good performance benchmark for QAOA. In Sec.~\ref{sec:numerics}, numerical simulations revealed that QAOA with fixed level $p$ is able to beat the greedy algorithm on average. In this section, we strengthen this result with numerical insights in the infinite size limit, $n\rightarrow\infty$.

To understand the performance of the greedy algorithm it is instructive to translate the action of the greedy algorithm presented in Fig.~\ref{alg:greedy} into the spin glass picture. For the sake of clarity of presentation, we assume that all $2n$ couplings have magnitude $|J|=1$, which is true in the infinite size limit. The greedy algorithm starts with assigning a random configuration to the spin representing the first car of the sequence. It then successively visits the spins representing the next car in the sequence. If it visits a car for the first time, it fixes the state of its spin such that the coupling between the car and its predecessor is fulfilled, i.e. same state for ferromagnetic- and opposite state for antiferromagnetic coupling.  

The greedy algorithm is guaranteed to satisfy a coupling every time it approaches a spin representing a car that has not been visited yet. This happens $n-1$ times. The remaining $2n-n+1$ connections in the full graph are, however, not taken into account. On average, the energy of these unseen connections is equal to zero. In total, for $n\rightarrow\infty$, the greedy algorithm generates a solution with average energy of $E_\mathrm{G}/n=-1$, which results in solutions with color changes growing according to $\mathbb{E}_\mathrm{G}(\Delta_C)=n/2$. 

In comparison, in the limit of $n\rightarrow\infty$, the reverse causal cones of the two-point correlation functions after $p$ levels of QAOA only include qubits in graphs which resemble trees of degree $4$ and coupling strengths $J=\pm 1$ (see Appendix~\ref{app:coupling_strengs}). Thus, for systems of infinite size, the expectation value of each two-local operator is the same on average and given by the expectation value calculated on a tree.

To calculate the expectation value with a state-vector simulation, we would have to include $3^{p+1}-1$ qubits, which is difficult to calculate classically even for shallow versions of QAOA. In \cite{Streif_2020}, the authors developed a method that substantially increases the simulation capabilities using the small tree width of the involved tensor networks \cite{markov2008simulating}. To calculate the two-point correlation functions on tree subgraph support, this method only scales polynomially in the number of qubits, but exponentially in the number of QAOA blocks. This allows us to calculate the expectation value up to $p=7$ levels of QAOA, including $6560$ qubits. With that we find average energies and color changes given in Tab.~\ref{tab:energies}. We note that the tensor network calculation also requires the optimization of the QAOA parameters. As the optimization might have found a local optimum rather than the global optimum, the values in Tab.~\ref{tab:energies} for QAOA represent lower bounds on the performance. If one would be able to find a better set of parameters, the performance of the algorithm could be even improved. 

\begin{figure}[t!]
    \centering
    \includegraphics[width=1\linewidth]{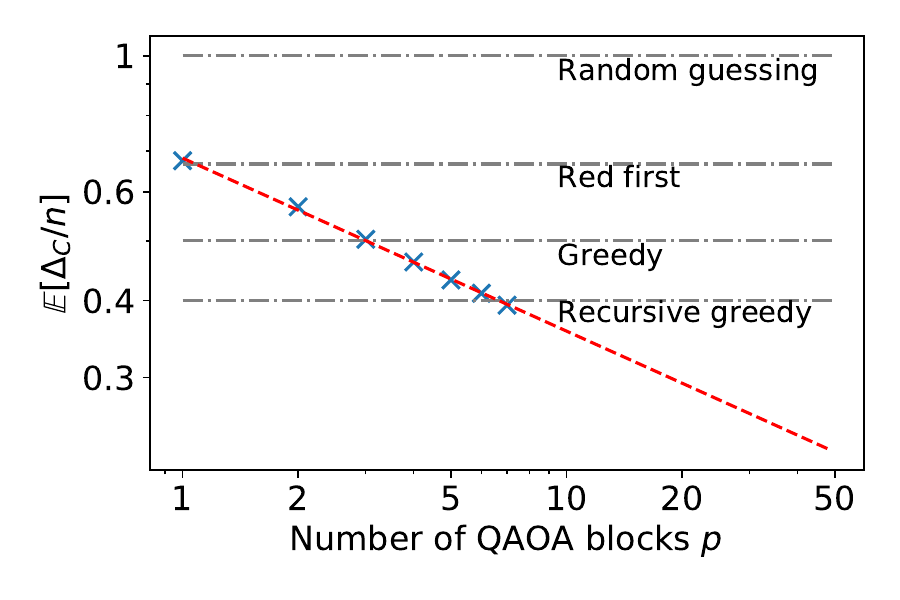}
  \caption{The data from Tab.~\ref{tab:energies} shown in a log-log plot together with a fit to the function $f(p)=10^b p^a$. The fit parameters are $a=(-0.279\pm 0.005)$ and $b=(-0.168\pm 0.003)$ with a coefficient of determination of $R^2=0.999$ \cite{draper1998applied}.}
  \label{fig:linear_regression_performance}
\end{figure}

We recognize that the performance of QAOA with $p=3$ levels is close to the performance of the greedy algorithm, while for $p=4$ there is a clear performance gap in favor of QAOA. While these arguments strictly hold for the limit of $n\rightarrow\infty$, we have shown that the results on smaller systems (see Fig.~\ref{fig:numerics}) agree with these results. When comparing the performance to two other heuristics, the red-first algorithm and the recursive greedy algorithm (see Appendix~\ref{app:classicalheuristics}), we see that QAOA outperforms the red-first algorithm on average with $p=2$ levels and for $p=7$ QAOA also beats the recursive greedy algorithm on average. In Fig.~\ref{fig:linear_regression_performance}, we show the data plotted against the number of QAOA blocks $p$ together with a fit highlighting the improvement in performance when increasing $p$.

To determine the run time of the quantum algorithm on a state-of-the-art quantum processing unit we recall that the binary paint shop problem can be represented as a graph of maximal degree 4 (see Sec.~\ref{sec:properties_spins}). On a fully-connected quantum computer, the corresponding QAOA circuit with a fixed value of $p$ thus only requires constant depth, making the QAOA for the BPSP a constant-time algorithm. This is in contrast with classical algorithms, which are polynomial-time algorithms. On quantum hardware with limited topology, e.g. for superconducting qubit devices with planar connectivity graphs, one could use methods such as the LHZ-encoding \cite{lechner2018quantum}, which require a quadratic overhead in qubits, however still have constant computing time.

\begin{table}[t!]
    \centering
 \begin{tabular}{||c | c ||} 
 \hline
 Method  & $\mathbb{E}\left[\Delta_{C}/n \right]$\\  
 \hline \hline
     Random guessing & 1.000\\
    \hline
 p=1 & 0.675 \\ 
 \hline
   Red first algorithm & 0.666\\  
 \hline
 p=2  & 0.568  \\
 \hline
 p=3  &0.503 \\
 \hline
  Greedy algorithm, see Fig.~\ref{alg:greedy}  &0.500 \\ 
   \hline
 p=4 &0.462  \\
 \hline
 p=5  & 0.432\\ 
  \hline
 p=6  & 0.411\\ 
  \hline
    Recursive greedy algorithm  & 0.400\\ 
   \hline
 p=7 & 0.393\\ 
 \hline

\end{tabular}
 \caption{Average performance in terms color changes of QAOA with different levels $p$ in comparison to greedy algorithms and random guessing in the limit $n\rightarrow\infty$. Ordered by the average performance.}
\label{tab:energies}
\end{table}

\subsection{Experimental results}
In this section, we show the results from QAOA circuits of binary paint shop instances with $p=1$ executed on a trapped-ion quantum computer, the IonQ device \cite{wright2019benchmarking}, provided by Amazon Braket \cite{AWS}. 
This device is composed of $11$ fully-connected qubits with average single- and two-qubit fidelities of $99.5\%$ and $97.5\%$ respectively \cite{wright2019benchmarking}. 
Like most available quantum hardware, trapped ion quantum computers only allow the application of gates from a restricted \textit{native gate set} predetermined by the physical realization of the processor. To execute an arbitrary gate, compilation of the desired gate into available gates is required. For trapped ions, a generic native gate set consists of a parameterized two-qubit rotation, $R_\mathrm{XX}(\alpha)=\mathrm{exp}[-\mathrm{i}\alpha \sigma_\mathrm{x}^{(i)}\sigma_\mathrm{x}^{(j)}/2]$ on qubits $i$ and $j$, and a single qubit rotation, $R$,
\begin{align}
    R(\theta,\phi)=
    \begin{pmatrix}
    \cos{(\theta/2)} & -\mathrm{i}\mathrm{e}^{-\mathrm{i}\phi}\sin{(\theta/2)} \\
    -\mathrm{i}\mathrm{e}^{\mathrm{i}\phi}\sin{(\theta/2)} &  \cos{(\theta/2)}  
    \end{pmatrix}
\end{align}
which includes $R_\mathrm{X}(\theta)=\mathrm{exp}[-\mathrm{i}\theta/2 \sigma_x^{(i)}]=R(\theta,0)$ and $R_\mathrm{Y}(\theta)=\mathrm{exp}[-\mathrm{i}\theta/2 \sigma_y^{(i)}]=R(\theta,\pi/2)$ \cite{maslov2017basic}. These gates form a universal set of gates, i.e. all other gates can be synthesized with these gates. 

The QAOA circuit, defined in Eq.~(\ref{eq:qaoa_ansatz}), includes the parameterized two-qubit rotation $R_\mathrm{ZZ}(\gamma)=\mathrm{exp}[-i\gamma \sigma_\mathrm{z}^{(i)}\sigma_\mathrm{z}^{(j)}]$ on qubits $i$ and $j$, parameterized single qubit $R_\mathrm{X}(\beta)$ rotations and Hadamard gates. While the local $R_\mathrm{X}(\beta)$ is readily available on the hardware and can be executed without any overhead, the Hadamard gate and the two-qubit $R_\mathrm{ZZ}(\gamma)$ rotation require compilation which will in turn increase the circuit depth. 

To make the circuit as short as possible, we rotate the circuit by inserting Hadamard gates. For the sake of clarity, we introduce the unitary $U_\mathrm{ZZ}=U_\mathrm{P}$, which highlights that the problem unitary is a set of $R_\mathrm{ZZ}$ gates. Accordingly, $U_\mathrm{XX}$ is the unitary in which all $R_\mathrm{ZZ}$ in $U_\mathrm{ZZ}$ were replaced by $R_\mathrm{XX}$ gates. The same pattern applies for the definition of the mixing unitary $U_\mathrm{X}$, i.e. $U_\mathrm{Y}$ denotes an unitary where all $R_\mathrm{X}$ gates are replaced by $R_\mathrm{Y}$ gates.  The new circuit,
\begin{align}
\label{eq:compiled_circuit}
    \ket{\Psi}_\mathrm{QAOA}^{p=1}&= U_\mathrm{X}(\beta) U_\mathrm{ZZ}(\gamma)\ket{+}\nonumber\\
    &=U_\mathrm{X}(\beta)  \mathrm{H}^\dagger \mathrm{H} U_\mathrm{ZZ}(\gamma)\mathrm{H}\ket{0}\nonumber\\
    &=U_\mathrm{X}(\beta)  \mathrm{H} U_\mathrm{XX}(\gamma)\ket{0}\nonumber\\
    &=U_\mathrm{X}(\beta-\pi)  U_\mathrm{Y}(\pi/2) U_\mathrm{XX}(\gamma)\ket{0}\;,
\end{align}
then only contains gates from the native gate set and thus needs no further compilation. For higher $p$-value, the transformation is analogous and shown in Appendix~\ref{app:circuit_optimization}.

On IonQ devices, all gates are executed in sequence \cite{ionq-website}. Thus, this representation of a QAOA circuit of a binary paint shop instance with $n$ nodes and $m$ edges can be carried out with circuit depth $d=m+2n$ and requires $2n$ single qubit gates and $m$ two-qubit gates. As the binary paint shop instances are bounded degree graphs with maximal degree 4, cf. Sec.~\ref{sec:properties_spins}, the circuit depth $d$ scales linearly with the system size $n$, $\mathcal{O}(d)\sim n$.

\begin{figure}[t!]
    \centering
    \includegraphics[width=1\linewidth]{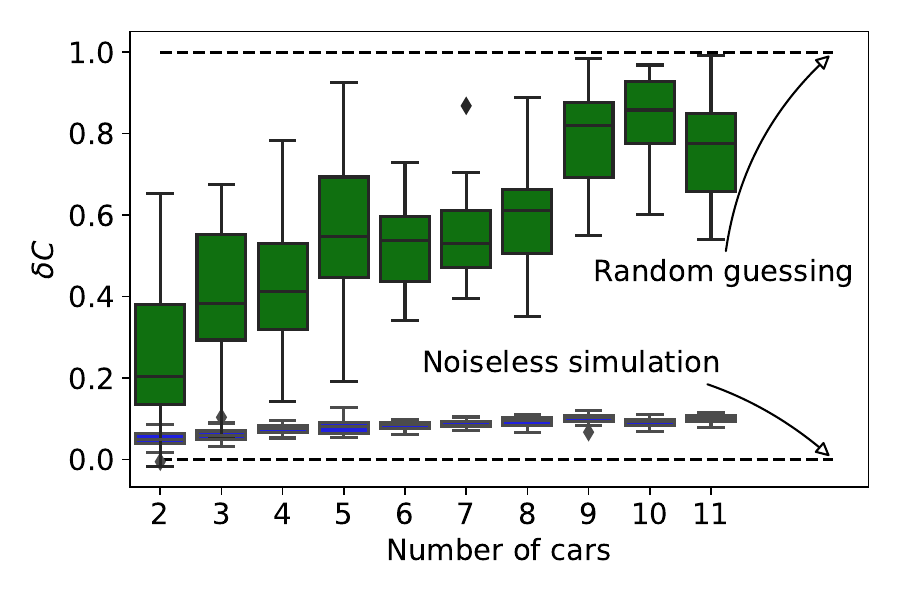}
  \caption{Performance of the QAOA experiments using an IonQ QPU (green) in comparison to a simulation with noise (blue), cf. Appendix~\ref{app:noise}, with $\delta C=\{0,1\}$ corresponding to random guessing and an ideal (noiseless) simulation respectively, as function of the number of cars. Results are presented at each system size for $N=20$ randomly drawn instances, averaged over $M=10^5$ measurements. In this box plot, the black line shows to the median, the green boxes the interquartile range (IQR), the whiskers $1.5$ times the IQR and the diamonds outliers. }
  \label{fig:performance_ionq}
\end{figure}

We execute the QAOA circuit with $p=1$ for $N=20$ randomly drawn binary paint shop instances from $n=2$ to $n=11$ cars \cite{bpsp_data}. For each instance, we take $M=10^5$ samples and calculate the average number of color changes, $\braket{\Delta_C^\mathrm{QPU}}$. For comparison, we take data from an ideal (noiseless) simulation and random guessing. To compare the experimental output with the ideal simulation and random guessing, we calculate the deviation in performance as 
\begin{align}
    \delta{C} = \frac{\braket{\Delta_C^\mathrm{QPU}}-\braket{\Delta_C^\mathrm{sim}}}{\braket{\Delta_C^\mathrm{random}}-\braket{\Delta_C^\mathrm{sim}}}
    \label{eq:performance_gap}
\end{align}
where $\braket{\Delta_C^\mathrm{sim}}$, $\braket{\Delta_C^\mathrm{QPU}}$ and  $\braket{\Delta_C^\mathrm{random}}$ denote the expected instance-wise color change obtained from the simulation, the QPU and random guessing respectively. A value of $\delta{C}=0$ implies that the QPU found results as good as the ideal simulation did, while $\delta{C}=1$ means that the QPU output mimics random guessing. 

In Fig.~\ref{fig:performance_ionq}, we plot the distribution of $\delta C$ over all $N$ instances for increasing system size. As clearly visible, for the smallest system size ($n=2$) the results are close to an ideal simulation, while for the largest studied instances ($n=11$), the output is almost random. Similar to previous QAOA experiments \cite{arute2020quantum, pagano2020quantum, otterbach2017unsupervised}, the present results highlight the strong influence of noise on the performance of the quantum algorithm. Moreover, in the same figure, we show the quantity $\delta C$ where we replaced the QPU results with obtained from a simulation with noise. In this simulation, the noise was tuned such that we find agreement with the reported single- and two-qubit fidelities in \cite{wright2019benchmarking}. For more details on the noise model used, we refer to Appendix~\ref{app:noise}. As visible, the simulation with noise cannot fully explain the experimental data. This could have several reasons:
\begin{enumerate}
    \item On IonQ all gates are carried out strictly successively \cite{ionq-website}, which leads to long idle times of the qubits. This is not taken into account by errors of single- and two-qubit gates. 
    \item The fidelites reported in \cite{wright2019benchmarking} were obtained for isolated 1- and 2-qubit gates.
\end{enumerate}
This outcome highlights that single- and two-qubit fidelities are not sufficient to fully characteristic the hardware's performance.

\section{Conclusion \& Outlook}
In this work, we applied QAOA to the binary paint shop problem, a computational problem from the automotive industry. We have shown numerical simulations together with experimental data obtained from a trapped ion quantum computer. Moreover, we were able to provide a comparison between the performance of QAOA and classical heuristics in the infinite size limit for noiseless quantum computation. 

The experimental results of this paper highlight the deterioration of the quantum algorithm's performance when increasing the problem size. To push forward to industry-relevant binary paint shop instances with hundreds of cars, either noise mitigation techniques or adaptions of QAOA must be developed to make this application on NISQ devices superior to random guessing. In this direction, the recursive adaption of QAOA introduced in \cite{bravyi2019obstacles} or the encoding of QAOA into a lattice gauge model \cite{lechner2018quantum} might provide improvements. Moreover, it would be interesting to investigate whether classical local algorithms \cite{hastings2019classical} are able to outperform the here shown results and to improve the classical performance bounds.

Furthermore, providing an answer on the question whether QAOA is a constant factor approximation algorithm could open up new room for quantum advantage.

This work has impact far beyond the restricted use case of the binary paint shop. In the way the binary paint shop problem is formulated we can already see that it is more general: every re-sequencing problem with a cost function that only depends on the  relative orientation of variables that are adjacent in the sequence is similar to the binary paint shop problem. Even a generalization to next-to adjacent orientation is straightforward. These kind of re-sequencing problems are prevalent in many production facilities. For example the optimisation problem of cars being assembled by workers: cars with sunroofs are more difficult to assemble and should not be adjacent in a sequence so that the workers can keep up with the constant speed of the conveyor belt. Furthermore, since we use a generic quantum  algorithm, QAOA, we are confident that the observed runtime and solution quality enhancements will prevail as long as we do not change the general problem characteristics: regular problem graph with interaction strengths of magnitude 1. These strict requirements can also be relaxed to a considerable amount as shown in \cite{Streif_2020}. We relegate the search for interesting use-cases within this group to future work.

\label{sec:conclusion}
\begin{acknowledgments}
This project has received funding from the European Union’s Horizon 2020 research and innovation programme under the Grant Agreement No. 828826. We thank Martin Schütz for technical support with Amazon Braket and H{\'e}ctor Valverde for technical support with AWS.
\end{acknowledgments}
\bibliography{lib}
\onecolumngrid
\appendix
\section{Classical heuristics}
\label{app:classicalheuristics}
\subsection{Red-first heuristic}
The red-first heuristic is a greedy algorithm for the binary paint shop problem in which all first occurrences of each car have the same color. This heuristic has proven average performance for $n\rightarrow\infty$ of
\begin{align}
    \lim_{n\rightarrow\infty}\mathbb{E}_\mathrm{RF}(\Delta_C)=\frac{2n}{3}\;.
\end{align}
After mapping the binary paint shop problem to an Ising Hamiltonian, the action of the red-first heuristic on the sequence representing a BPSP instance is equivalent to setting all spins in the spin system to the same value. The average energy for $n\rightarrow\infty$ is given by 
\begin{align}
    \lim_{n\rightarrow\infty}\mathbb{E}_\mathrm{RF}(E)=\frac{1}{2}\sum J_{ij} s_i s_j = \frac{1}{2} \sum J_{ij}=-\frac{n}{3}\;,
\end{align}
where we used the results from Appendix~\ref{app:coupling_strengs} on the average degree and coupling strengths of the graph.

\subsection{Recursive greedy heuristic}

The recursive greedy heuristic starts by iteratively deleting both occurrences of the last car of the sequence until the sequence has length $2$ and only one car. Subsequently, the sequence of length $2$ is painted optimally. After that, the occurrences of the last deleted car are added back to the sequence. While keeping the already painted cars fixed, the new car is painted optimally. This is repeated until the whole sequence is painted.

From the spin system perspective, this corresponds to the procedure of deleting all spins expect a single spin and adding back spin by spin while setting the each state to the best possible energetic configuration. 

This heuristic has a proven average performance for $n\rightarrow\infty$ of
\begin{align}
    \lim_{n\rightarrow\infty}\mathbb{E}_\mathrm{RG}(\Delta_C)=\frac{2n}{5}\;.
\end{align}

As single color change yields an increase of energy by $1$, we use the results on the red-first heuristic to determine the average energy to
\begin{align}
    \lim_{n\rightarrow\infty}\mathbb{E}_\mathrm{RG}(E)=-\frac{3n}{5}\;.
\end{align}

\subsection{Random guessing}
As a baseline for comparison we here show the performance of random guessing. For $n\rightarrow\infty$,
\begin{align}
    \lim_{n\rightarrow\infty}\mathbb{E}_\mathrm{random}(E)=\sum_{s_i,s_j=-1,1} \sum_{i,j}J_{ij}s_is_j = \sum_{i,j} (1/2 J_{ij}-1/2 J_{ij}) = 0\,,
\end{align}
and thus 
\begin{align}
    \lim_{n\rightarrow\infty}\mathbb{E}_\mathrm{random}(\Delta_C)=n\;.
\end{align}
\section{Characteristics of the spin system}
\label{app:coupling_strengs}
In this section of the Appendix, we discuss the distribution of coupling strengths $J_{ij}$ and the average degree of the Ising model resulting from the mapping given in Fig.~\ref{alg:spin_glass_mapping}.

\paragraph{Coupling strengths}
We here show that the probability of finding a ferromagnetic interaction ($J=-1$) in the spin glass representation of the binary paint shop is twice as big as finding an anti-ferromagnetic coupling ($J=1$) when $n\rightarrow\infty$, while the probability of finding a coupling of $J=-2$ is converging to zero.

\begin{figure*}[t!]
    \centering
    \begin{minipage}{0.5\textwidth}
        \centering
        \includegraphics[width=1\textwidth]{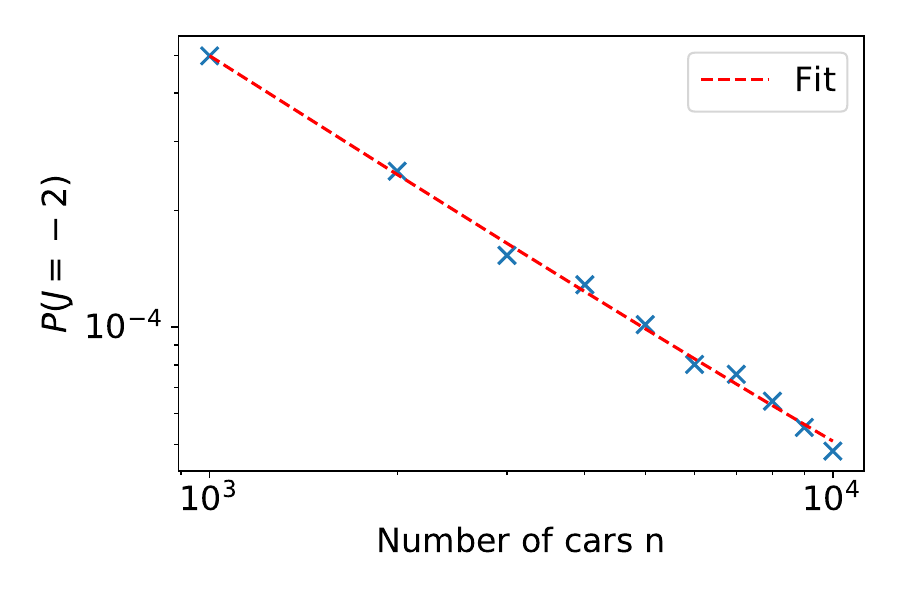}\llap{
  \parbox[b]{6.7in}{(a)\\\rule{0ex}{2.1in}
  }}
    \end{minipage}\hfill
    \begin{minipage}{0.5\textwidth}
        \centering
        \includegraphics[width=1\textwidth]{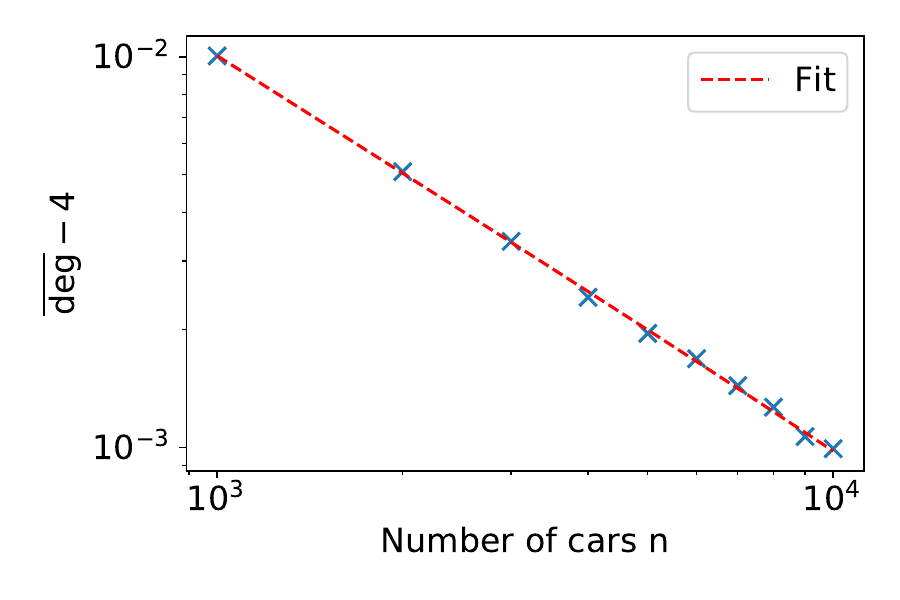}\llap{ \parbox[b]{6.7in}{(b)\\\rule{0ex}{2.1in}
  }}\par
\vskip\floatsep%
    \end{minipage}\hfill
    \centering
    \includegraphics[width=0.5\textwidth]{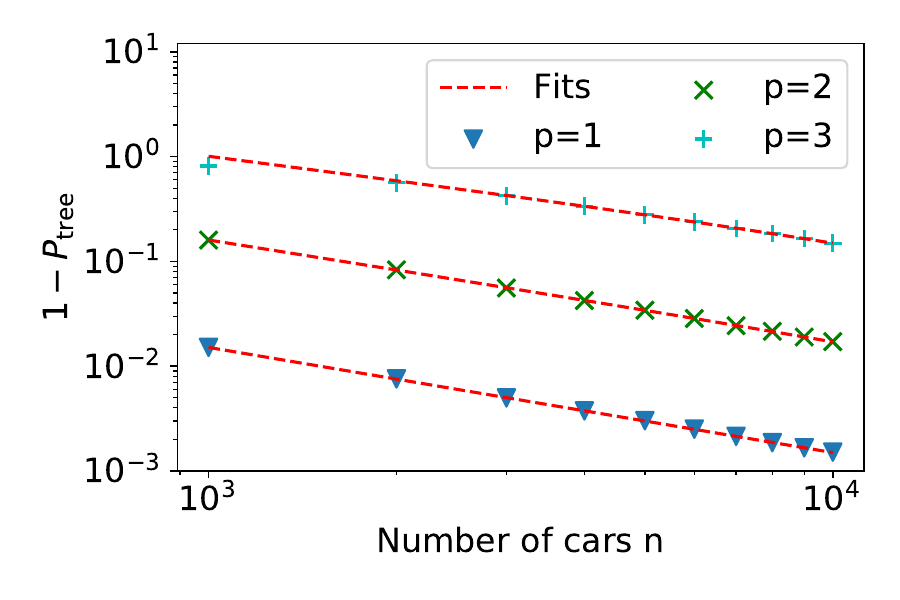}\llap{ \parbox[b]{7in}{(c)\\\rule{0ex}{2.1in}
  }}
\caption{This plot provides numerical insight into the properties of the graph representation of the binary paint shop problems for increasing system size $n$. For all plots, we fit the data to a function of the form $f(n)=10^a n^b+c$. (a) The probability to draw an edge with coupling strength $J=\{-2\}$. Fit parameters are $(b=-1.028\pm 0.048, c=0.000\pm 0.000)$ (b) The deviation of the average degree $\overline{\mathrm{deg}}$ of random binary paint shop instances from degree 4. Fit parameters are $(b=-0.991\pm0.019, c=0.000\pm 0.000)$ (c) The probability to draw a QAOA subgraph which is not a tree for $p=1,2,3$ levels of QAOA. Fit parameters are $(b=-1.006\pm0.006, c=9\cdot 10^{-6}\pm 3\cdot10^{-5})$ for $p=1$, $(b=-0.980\pm0.005, c=-0.002\pm 0.000)$ for $p=2$ and $(b=-0.732\pm0.029 , c=-0.046\pm 0.010)$ for $p=3$. For the fit of $p=3$ only the last six data points were used. All data points were averaged over $1000$ randomly drawn instances for each system size.}
\label{fig:coupling_prob_and_average_degree}
\end{figure*}

If we draw a random pair of cars $(w_i,w_{i+1})$ of the sequence, the probability that we find twice the same car $c_i$ is given by $P_\mathrm{same car}=1/(2n-1)$. A similar argument can be made for the probability that a single car $c_i$ has the same neighbor $c_j$ twice. As couplings of $J=-2$ are only generated when a car has the same neighbor twice, this already means that $P_{ij}(-2)=0$ when $n\rightarrow\infty$.  This is in agreement with the numerics shown in Fig.~\ref{fig:coupling_prob_and_average_degree}(a).

Thus, when looking at $n\rightarrow\infty$, we now exclude the cases of the previous paragraph and look at the probabilities $P_{ij}(+1)$ and $P_{ij}(-1)$. By construction, a ferromagnetic coupling is generated whenever two cars in the sequence are neighbors and both occur for the first time or both already have occurred. If we draw a random pair of the sequence, the probability that this pair generates a ferromagnetic coupling is given by
\begin{align}
    P(J=-1)&=\frac{1}{2n-1}\sum_{<ij>}\left[P_{ij}(00)+P_{ij}(11)\right]\;,
    \label{eq:probability_J}
\end{align}
where $P_{ij}(00)$ is the probability that both cars at positions $i$ and $j$ in the sequence occur the first time, and $P_{ij}(11)$ that both cars appear for the second time. For position $i$ in the sequence, the probability that the car already appeared before is given by $(i-1)/2n$. With this, we reformulate Eq.~(\ref{eq:probability_J}) as
\begin{align}P(J=-1)=\frac{1}{2n-1}\sum_{i=1}^{2n-1}\left[\left(1-\frac{i-1}{2n-1}\right)\left(1-\frac{(i+1)-1}{2n-1}\right)+\frac{i-1}{2n-1}\frac{(i+1)-1}{2n-1}\right]=\frac{1-6n+8n^2}{6n(2n-1)}
\end{align}
For $n\rightarrow\infty$, we find $P(J=-1)=2/3$, i.e. the probability of finding a ferromagnetic interaction strength is $2/3$.
The probability of finding an anti-ferromagnetic coupling, $P(J=+1)$, can be calculated in a similar way by calculating the probability that one of two consecutive cars in the sequence was already seen while the other did not occur yet. This can be written as
\begin{align}
    P(J=+1)&=\frac{1}{2n-1}\sum_{<ij>}\left[P_{ij}(01)+P_{ij}(10)\right]\\
    &=\frac{1}{2n-1}\sum_{i=1}^{2n-1}\left[\left(\frac{i-1}{2n-1}\right)\left(1-\frac{(i+1)-1}{2n-1}\right)+\frac{i-1}{2n-1}\left(1-\frac{(i+1)-1}{2n-1}\right)\right]=\frac{4n^2 -1}{6n(2n-1)},
\end{align}
which for $n\rightarrow\infty$ gives $P_{ij}(J=+1)=1/3$.
Thus, the spin system formulation of the binary paint shop problems have (for $n\rightarrow\infty$) integer ferromagnetic or anti-ferromagnetic couplings with probabilities $2/3$ and $1/3$ respectively.

\paragraph{Average degree}
As the probability that a car is twice next to the same car in the sequence is vanishing when $n\rightarrow\infty$, each car representing a spin in the spin system has 4 connections in the infinite size limit. This intuition is supported by Fig.~\ref{fig:coupling_prob_and_average_degree}(b), where we show the deviation from an average degree of $4$ for $1000$ randomly drawn binary paint shop instances while increasing the system size. 

\paragraph{Number of tree subgraphs}
The expectation value of QAOA is given by the sum of the expectation values of two-point correlation functions $\sigma_z^{(i)}\sigma_z^{(j)}$, see Eq.~\ref{eq:sum_expectation_value}. As discusses in Sec.~\ref{sec:solving_the_bpsp_with_qaoa}, each expectation value can be calculated over the reverse causal cone of the corresponding edge $(i,j)$ in the graph. In Fig.~\ref{fig:coupling_prob_and_average_degree}(c) we show the probability $1-P_\mathrm{tree}$ with $P_\mathrm{tree}$ the probability that a randomly drawn subgraph resembles a tree of degree 4 for $p=1$, $p=2$ and $p=3$. The numerical experiment together with the fits suggest that the probability to draw subgraphs which are not tress is zero in the infinite size limit ($n\rightarrow\infty$).

\section{Extension of tree-QAOA}
\label{app:tree-qaoa}
In this section, we prove that the optimal QAOA parameters on a tree-graph with coupling strength $J=+1$ are equivalent to the optimal QAOA parameters on a tree-graph with coupling strengths $J=\pm 1$.
As a starting point, we assume that we have an Ising model defined on a tree,
\begin{align}
    H_\mathrm{tree} = \sum_{(i,j)\in E} J_{ij} \sigma_i \sigma_j\;,
\end{align}
with the edge set $E$ defining the tree graph. 
To prove the assumption, we show that the optimal parameters of tree-QAOA stay the same when replacing the coupling strengths $J_{ij}=1$ with any $J_{ij}=\pm 1$. We assume that there exist a transformation
\begin{align}
    \label{eq:trafo}
    U_\mathrm{T} H^{J=1}_\mathrm{tree} U_\mathrm{T}^\dagger = H^{J=\pm1}_\mathrm{tree}, 
\end{align}
with $U_\mathrm{T}=\bigotimes_{j}^k X_j$ defining a $k$-local spin flip operation on a subset of $k$ qubits. Inserting this transformation into the expectation value of the QAOA circuit, Eq.~(\ref{eq:expectation_value_qaoa}), yields
\begin{align}
     &\braket{+|\dots \mathrm{e}^{i\gamma_p H^{J\pm 1}_\mathrm{tree} }U_\mathrm{X} (\pm Z_i Z_j)U^\dagger_\mathrm{X}\mathrm{e}^{-i\gamma_p H^{J\pm 1}_\mathrm{tree} } \dots|+}\nonumber\\
    =&\braket{+|\dots \mathrm{e}^{i\gamma_p U_\mathrm{T} H^{J=1}_\mathrm{tree} U_\mathrm{T}^\dagger}U_\mathrm{X} (\pm Z_i Z_j) U^\dagger_\mathrm{X}\mathrm{e}^{-i\gamma_p U_\mathrm{T} H^{J=1}_\mathrm{tree} U_\mathrm{T}^\dagger}\dots|+}\nonumber\\
    =&\braket{+|\dots U_\mathrm{T} \mathrm{e}^{i\gamma_p H^{J=1}_\mathrm{tree} } U_\mathrm{T}^\dagger U_\mathrm{X} (\pm Z_i Z_j)U_\mathrm{X} U_\mathrm{T}\mathrm{e}^{-i\gamma_p H^{J=1}_\mathrm{tree} } U_\mathrm{T}^\dagger\dots|+}\nonumber\\
    =&\braket{+|\dots \mathrm{e}^{i\gamma_p H^{J=1}_\mathrm{tree} } U_\mathrm{T}^\dagger U_\mathrm{X} (\pm Z_i Z_j) U^\dagger_\mathrm{X}U_\mathrm{T}\mathrm{e}^{-i\gamma_p H^{J=1}_\mathrm{tree} } \dots|+}\nonumber\\
    =&\braket{+|\dots \mathrm{e}^{i\gamma_p H^{J=1}_\mathrm{tree} } U_\mathrm{X} U_\mathrm{T}^\dagger (\pm Z_i Z_j)U_\mathrm{T} U^\dagger_\mathrm{X}\mathrm{e}^{-i\gamma_p H^{J=1}_\mathrm{tree} }\dots|+}\nonumber\\
    =&\braket{+|\dots\mathrm{e}^{i\gamma_p H^{J=1}_\mathrm{tree} } U_\mathrm{X}(Z_i Z_j)U^\dagger_\mathrm{X}\mathrm{e}^{-i\gamma_p H^{J=1}_\mathrm{tree} } \dots|+}
\end{align}
This shows that, if a transformation Eq.~(\ref{eq:trafo}) exists, then the expectation value of the tree-QAOA with couplings strengths $J_{ij}=\pm1$ is equivalent to the initial case with $J_{ij}=1$ and the variational parameters are the same. On trees, where no frustration is present, it is always possible to find such a transformation.

\section{Circuit optimization for trapped ion quantum computers}
\label{app:circuit_optimization}
In Eq.~(\ref{eq:compiled_circuit}), we showed how to transform the QAOA with $p=1$ circuit such that only native gates are used. In this section, we show how this can be done for an arbitrary number of QAOA levels $p$. By inserting Hadamard gates, the circuits transforms to 
\begin{align}
    \ket{\Psi}_\mathrm{QAOA}^{p} &= U_\mathrm{X}(\beta_p) U_\mathrm{ZZ}(\gamma_p)\dots U_\mathrm{X}(\beta_2) U_\mathrm{ZZ}(\gamma_2) U_\mathrm{X}(\beta_1) U_\mathrm{ZZ}(\gamma_1)\ket{+}\nonumber\\
    &= U_\mathrm{X}(\beta_p) \mathrm{H}\mathrm{H} U_\mathrm{ZZ}(\gamma_p)\mathrm{H}\mathrm{H}\dots \mathrm{H}\mathrm{H}U_\mathrm{X}(\beta_2) \mathrm{H}\mathrm{H}U_\mathrm{ZZ}(\gamma_2)\mathrm{H}\mathrm{H} U_\mathrm{X}(\beta_1)\mathrm{H}\mathrm{H} U_\mathrm{ZZ}(\gamma_1)\mathrm{H}\ket{0}\nonumber\\
    &= U_\mathrm{X}(\beta_p) \mathrm{H} U_\mathrm{XX}(\gamma_p)\dots U_\mathrm{Z}(\beta_2)  U_\mathrm{XX}(\gamma_2) U_\mathrm{Z}(\beta_1) U_\mathrm{XX}(\gamma_1)\ket{0}\nonumber\\
    &= U_\mathrm{X}(\beta_p-\pi) U_\mathrm{Y}(\pi/2) U_\mathrm{XX}(\gamma_p)\dots U_\mathrm{Z}(\beta_2)  U_\mathrm{XX}(\gamma_2) U_\mathrm{Z}(\beta_1) U_\mathrm{XX}(\gamma_1)\ket{0}\;,
\end{align}
including only native gates.

\section{Noise model}
\label{app:noise}
In this section of the Appendix we introduce the noise model used in the noisy simulation shown in Fig.~\ref{fig:performance_ionq}. After each application of a gate on a set of qubits, we subsequently apply a local depolarizing channel,
\begin{align}
\mathcal{E}(\rho):\rho \rightarrow \sum_i K_i\rho K_i^\dagger
\end{align}
with the Kraus operators
\begin{align}
   K_1&= \sqrt{1-\eta}\mathbb{1},\hspace{0.3cm} &K_2&= \sqrt{\frac{\eta}{3}}\sigma_x,\hspace{0.3cm} & K_3&=\sqrt{\frac{\eta}{3}}\sigma_y,\hspace{0.3cm} &K_4&=\sqrt{\frac{\eta}{3}}\sigma_z,
\end{align}
on all qubits which participated in the gate. Thus, for a single-qubit gate on qubit $i$, we apply a single local depolarizing channel on qubit $i$. For a two-qubit gate on qubits $i,j$, we apply two local depolarizing channels on qubits $i,j$. As two-qubit gates are more error-prone than one-qubit gates, we assign different error rates to single- and two-qubit gates, $\eta_{1\mathrm{Q}}$ and $\eta_{2\mathrm{Q}}$ respectively. On IonQ, the average single- and two-qubit fidelities are reported to be $99.5\%$ and $97.5\%$ respectively \cite{wright2019benchmarking}. To adjust the error rate of the single-qubit gate, $\eta_{1\mathrm{Q}}$, we follow the experimental procedure from \cite{wright2019benchmarking} and simulate random benchmarking and measure the fidelity. We tune the error rate $\eta_{1\mathrm{Q}}$ such that we find a fidelity of $99.5\%$, resulting in $\eta_{1\mathrm{Q}}=0.0029$ Similarly, to find an appropriate value for $\eta_{2\mathrm{Q}}$, we simulate partial state tomography on the Bell state and tune $\eta_{2\mathrm{Q}}$ resulting in $\eta_{2\mathrm{Q}}=0.0168$.

\begin{turnpage}
\begin{table}
\begin{tabular}{c||c|c|c|c|c|c|c|c|c|c|c|c|c|c|}
    $p$ & $\gamma_1$ & $\beta_1$ & $\gamma_2$ & $\beta_2$ & $\gamma_3$ & $\beta_3$ & $\gamma_4$ & $\beta_4$ & $\gamma_5$ & $\beta_5$ & $\gamma_6$ & $\beta_6$& $\gamma_7$ & $\beta_7$ \\ 
    \hline \hline
     1 & $ 0.52358$ & $-0.39269$ &  &  &  &  &  &  &  &  &  &  &  &  \\ 
     2 & $ 0.40784$ & $ -0.53411$ & $ 0.73974$ & $ -0.28296$ &  &  &  &  &  &  &  &  &  &  \\ 
     3 & $ 0.35450$ & $ -0.58794$ & $ 0.65138$ & $ -0.42318$ & $0.75426$ & $ -0.22301$ &  &  &  &  &  &  &  &  \\ 
     4 & $ 0.31500$ & $ -0.60498$ & $ 0.58754$ & $ -0.47780$ & $ 0.67322$ & $- 0.36127$ & $ 0.77120$ & $ -0.18753$ &  &  &  &  &  &  \\ 
     5 & $ 0.29092$ & $ -0.62254$ & $ 0.54678$ & $ -0.50507$ & $ 0.60334$ & $ -0.41672$ & $ 0.68722$ & $- 0.32534$ & $ 0.78446$ & $- 0.16280$  &  &  &  & \\ 
     6 & $ 0.26872$ & $ -0.62933$ & $ 0.51278$ & $ -0.52317$ & $ 0.56359$ & $ -0.45282$ & $ 0.61410$ & $- 0.38834$ & $ 0.69565$ & $- 0.29814$  & $0.78667$ & $-0.14595$& & \\ 
     7 & $ 0.25377$ & $ -0.63776$ & $ 0.48903$ & $ -0.53260$ & $ 0.53172$ & $ -0.47185$ & $ 0.57572$ & $- 0.43247$ & $ 0.62149$ & $- 0.36318$& $0.69778$& $- 0.27774$& $ 0.78866$ & $-0.13380$ \\ 
\end{tabular}
\caption{The QAOA parameters obtained with the method given in \cite{Streif_2020} and used for the numerical simulations and experiments in this paper.}
\label{tab:tree-qaoa-parameters}
\end{table}
\end{turnpage}
\end{document}